\documentclass[twocolumn, prl, aps, superscriptaddress, longbibliography,showpacs,amsmath,amssymb,floatfix]{revtex4-1}
\usepackage{changes}
\usepackage{graphicx}
\usepackage{dcolumn}
\usepackage{bm}
\usepackage{graphicx}                                   
\usepackage{amssymb}

\usepackage{amsmath}
\usepackage{epsfig}
\usepackage{xcolor}
\usepackage{tabu}
\usepackage{mathtools}
\usepackage[colorlinks,linkcolor=blue,anchorcolor=blue,citecolor=blue,urlcolor=blue]{hyperref}
\usepackage{physics}
\usepackage{float}
\usepackage{diagbox}
\usepackage{inputenc}

\begin{document}
    
\title{Effective diffusion of Brownian motion in spatially quasi-periodic noise}
\author{Sang Yang}
\email{yangsang@mail.ustc.edu.cn}
\affiliation{University of Science and Technology of China, Hefei, 230026, China}

\author{Zhixin Peng}
\affiliation{University of Science and Technology of China, Hefei, 230026, China}


\date{\today}
	
\begin{abstract}
The effective diffusion of Brownian particles in periodic potential has been a central topic in nonequilibrium statistical physcis. A classical result is the Lifson formula which provides the effective diffusion constant in periodic potentials. Extending beyong periodicity, our recent work [arXiv:2504.16527] has demonstrated that a modified Lifson expression remains valid for Brownian motion in quasi-periodic potentials. In this work, we extend our previous results by incorporating spatial quasi-periodic noise and examining different stochastic interpretations, $\alpha\in[0,1]$. The proposed framework is simple, computationally efficient, and unifies the treatment of diffusion in both periodic and quasi-periodic systems.
\end{abstract}
\maketitle
	
The Brownian motion describes the dynamics of a single particle undergoing random motion driven by stochastic forces \cite{Bian2016111Years, ChandrasekharReview, mel1991kramers, reimann2002brownian, 
hanggi2009artificial, hanggi1990reaction}, whose strength is related to the system temperature through Eisntein relation \cite{einstein1905molekularkinetischen}, or, more generally, the flucutation-dissipative theorem \cite{kubo1966fluctuation}. In the overdamped regime, the Langevin equation for Brownian motion in a periodic or quasi-periodic potential reads 
\begin{equation}
    \gamma\frac{\text{d}x}{\text{d}t}=-\frac{\text{d}U(x)}{\text{d}x}+\xi(t).
    \label{eqn_brown}
\end{equation}
Here $\gamma$ is the friction coefficient depending on the particle size, and $U(x)$ is the external potential. Here $x(t)$ represents the particle trajectory under the stochastic force $\xi(t)$, which satisfies
\begin{equation}
\langle\xi(t)\rangle =0, \quad \langle\xi(t) \xi(t')\rangle = 2k_BT\gamma\delta(t-t').
\end{equation}
When the external potential $U(x)$ is periodic, periodic diffusion model can be used to describe a wide range of phenomena in physics and related fields, such as current transport in Josephson junctions \cite{Ambegaokar1969Voltage, jack2017quantum, bishop1978josephson, coffey2009nonlinear, longobardi2011thermal, zwerger1987quantum}, surface atom diffusion \cite{speer2009directing, AlaNissila01052002, Lacasta2004Fromsubdiffusion, Sancho2004Diffusion}, thermal ratchets \cite{reimann2002brownian, hanggi2009artificial, Bier01111997, Renzoni01052005Cold, Julicher1997Modeling}, and particle dynamics in corrugated or confined channels \cite{Palastro2008Pulse, Yang2019Diffusion, Ghosh2012Brownian, Reguera2001Kinetic, Jeon2010Fractional}. Periodic models have also been applied to cold atoms \cite{Kindermann2017Nonergodic}, among many others. Despite these extensive applications, diffusion in quasi-periodic potentials—which, in fact, encompasses all periodic cases—has not yet been systematically investigated, either theoretically or experimentally, prior to our recent work [arXiv:2504.16527]. Assumming a more general form of the potential, $U(x)=U_a\sin(ax)+U_b\sin(bx)$, with an irrational ratio $a/b\notin \mathcal{Q}$. For $U_b=0$, the poetntial $U(x)$ is periodic, whereas for $U_b\neq 0$, it becomes quasi-periodic. In the absence of an external potential ($U_a=U_b=0$) and in homogeneous environments, the diffusion coefficient is given by Einstein relation $D_0=k_BT/\gamma$. For periodic potentials ($U_b=0$), the well-known Lifson formula accurately describes the long-time transport behavior,
\begin{equation}
    D^*=\frac{D_0}{\langle \exp(-\beta U)\rangle_{R_a}\langle \exp(\beta U)\rangle}_{R_a},
    \label{lifson}
\end{equation}
where $R_a=2\pi/a$ is the periodic length. For quasi-periodic potentials, our rencent work [arXiv:2504.16527] has shown that a modified Lifson expression remains valid,
\begin{equation}
    D^*=\frac{D_0}{\langle \exp(-\beta U)\rangle_{L}\langle \exp(\beta U)\rangle}_{L},
    \label{lifson-md}
\end{equation}
where $L$ being a characterized length scale. Moreover, we propose a unified and simplified form that encompasses all of the these cases. Remarkably, even for more general potentials of the type $U(x)=\sum_i U_i\sin(2\pi x/R_i)$, the effective diffusion coefficient can be expressed as $D^*=D_0/\Pi_iI_0(U_i)$, which depends only on poetential amplitudes, $U_i$.

In this paper, we extend the discussion to a more general setting where the noise is spatially dependent, with particluar emphasis on spatially quasi-periodic noise. As we all known, spatially periodic noise has been investigated extensively, most notably in the seminal works of Büttiker \cite{buttiker1987transport} and Landauer\cite{landauer1988motion}. Büttiker considered the interplay between a periodic drift and periodic noise, while Landauer examined systems with nonuniform temperature and demonstrated that particles tend to leave hot regions more rapidly than cold ones. More recently, Giordano and Blossey derived the effective diffusion coefficient for stochastic processes subject to spatially periodic noise under different stochastic interpretations\cite{giordano2024effective}. 

The role of stochastic interpretations becomes even more critical when considering spatially quasi-periodic noise. These interpretations originate from Itô's pioneering solution of Langevin equation with Itô calculus, later extended by Stratonovich, Hänggi, and collaborators, who introduced alternative schemes. Together, these works established the freedom in choosing the discretization point in stochastic integration. In practice, one introduces a real parameter $\alpha\in [0,1]$ to distinguish different schemes: the Itô, Stratonovich, and Hänggi integrations correspond to $\alpha=0,1/2,1$, respectively. While all three yield identical results in homogeneous systems ($\mathrm{d}D_0/\mathrm{d}x = 0$), they play a decisive role in heterogeneous cases, particularly when the diffusion coefficient $D(x)$ varies spatially. In such cases, the classical Lifson formula must be modified, 
\begin{equation}
    D^*=\frac{1}{\langle \exp(-\beta U)\rangle_{R_a}\langle \exp(\beta U)/D(x)\rangle}_{R_a},
    \label{weaver}
\end{equation}
as first demonstrated by Weaver [Eq. (9)–(10) in Ref. \cite{weaver1979effective,gunther1979mobility}]. Importantly, Weaver’s results can be rederived more transparently by recognizing the equivalence between a spatially varying diffusion coefficient and an effective potential, namely, by treating $\ln D(x)$ as an effective potential (see Appendix). This perspective forms one of the two approaches adopted in the present work. The second approach is based on the function separation method, in close analogy to the procedure we have recently developed for handling diffusion in quasi-periodic potentials [arXiv:2504.16527]. 

The article is organized as follows. In the first part, we introduce the effective potential method to evaluate the effective diffusion coefficient. In the second part, we present the function separation method, which extends our previous approach. Finally, in the Appendix, we demonstrate how the effective potential method provides a simplified derivation of Weaver’s results.

To proceed, we first write down the Smoluchowski equation for Brownian motion driven by quasi-periodic noise in a quasi-periodic potential,
\begin{equation}
    \frac{\partial p(x,t)}{\partial t} = \frac{\partial}{\partial x} D^{\alpha} e^{-\beta U}\frac{\partial}{\partial x} D^{1-\alpha} e^{-\beta U}p(x,t),
\end{equation}
where $ \alpha $ denotes the stochastic interpretation parameter, $U(x)$ is a quasi-periodic potential, and $D(x)$ is a quasi-periodic diffusion coefficient. The later may araise from the heterogeneity of the friction coefficient $\gamma(x)$ or the local temperature $T(x)$.

{\it The effective potential method.} We begin with the simplest case, in the absence of an external potential. In this case, the Smoluchowski equation reduces to
\begin{equation}
    \frac{\partial p(x,t)}{\partial t} = \frac{\partial}{\partial x} [D] \tilde{D}^{\alpha} \frac{\partial}{\partial x} \tilde{D}^{1-\alpha} p(x,t).
    \label{sm-withou-U}
\end{equation}
where, $[D]$ is the dimension, 1 m$^2$/s, which can keep $\tilde{D}=D/[D]$ be dimensionless. Using the transformation $\tilde{D}^{\alpha}=\tilde{D} \tilde{D}^{-(1-\alpha)} $, one obtain 
\begin{equation}
    \frac{\partial p(x,t)}{\partial t} = \frac{\partial}{\partial x} D(x) \tilde{D}^{-(1-\alpha)} \frac{\partial}{\partial x} \tilde{D}^{1-\alpha} p(x,t),
\end{equation}
which directly connects to the case with an external potential \cite{weaver1979effective,pacheco2024langevin}. By Noting the correspondence between $\tilde{D}^{-(1-\alpha)}$ and $\exp(-\beta \tilde{U}(x))$, the spatially varying diffusion coefficient can be mapped onto an effective potential $\tilde{U}$ such that
\begin{equation}
    \pm \beta \tilde{U} = (1\mp\alpha)\ln \tilde{D}.
    \label{D-tilde-without-U}
\end{equation}
Substituting this mapping into the known expression for the effective diffusion constant in heterogeneous environments \cite{weaver1979effective}, we obtain the final formula,
\begin{eqnarray}
    D^{\ast} &=&\frac{1}{\langle e^{-\beta \tilde{U}(x)}\rangle\langle e^{\beta \tilde{U}(x)}/D(x)\rangle} \nonumber\\
    &=&\frac{1}{\langle D^{-(1-\alpha)} \rangle_L \langle D^{-\alpha} \rangle_L}
    \label{Dstar-alpha-1}
\end{eqnarray}
In particular, for the Stratonovich interpretation ($\alpha=1/2$), this reduces to
\begin{equation}
    D^{\ast} = \frac{1}{\langle D^{-1/2} \rangle_L \langle D^{-1/2} \rangle_L} = (\langle D^{-1/2} \rangle_L)^{-2}.
\end{equation}
which agrees with the result of Giordano and Blossey \cite{giordano2024effective}, albeit derived there only for spatially periodic diffusion and periodic noise only. When an external potential $U(x)$ is present, the effective potential is modified as
\begin{equation}
    \pm \beta \tilde{U} = \pm\beta U(x)+(1\mp\alpha)\ln \tilde{D}.
    \label{D-tilde-with-U}
\end{equation}
and the corresponding effective diffusion coefficient takes the form
\begin{equation}
    D^{\ast} = \frac{1}{\langle \frac{e^{-\beta U(x)}}{D^{1-\alpha}} \rangle_L \langle \frac{e^{\beta U(x)}}{D^{\alpha}} \rangle_L}.
    \label{Dstar-alpha-2}
\end{equation}

\begin{figure}
\centering
\includegraphics[width=0.45\textwidth]{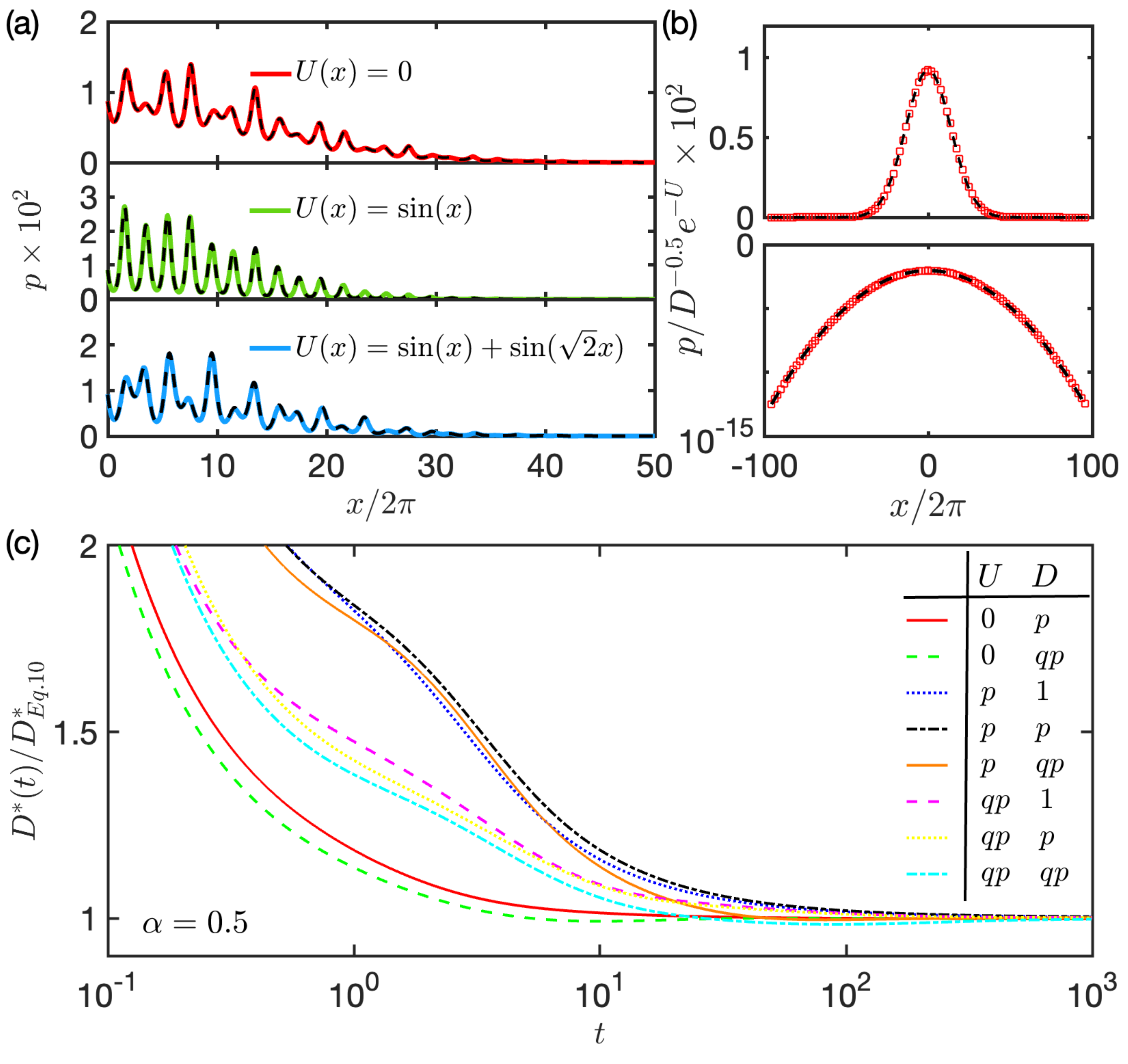}
\caption{The effective diffusion with different intepreter factor. (a) The probability density function $p(x,t)$ with $\alpha=1/2$ and $D(x)=1+\sin(x)/4+\sin(x/\sqrt{2})/4$. (b) Gussian-like function $g(x,t)$ with same parameter as (a). (c) The effective diffusion cosntant with different diffusion coefficients and potential functions (eqn. \ref{Dstar-alpha-1} and \ref{Dstar-alpha-2}). The periodic and quasi-periodic potential are $\sin(x)$ and $\sin(x)/2+\sin(\sqrt{3}x)/2$, respectively. The periodic and quasi-periodic diffusion coefficient are $1+\sin(x)/2$ and $1+\sin(x)/4+\sin(\sqrt{2})/4$, respectively.}
\label{fig-fig1}
\end{figure}

Whether or not an external potential is present, both transformations (Eq. \ref{D-tilde-without-U} and \ref{D-tilde-with-U}) recast the Smoluchowsk equation (Eq. \ref{sm-withou-U}) into the unified form
\begin{equation*}
	\frac{\partial p(x, t)}{\partial t}=\frac{\partial}{\partial x}\left\{D(x)e^{-\beta \tilde{U}(x)} \frac{\partial}{\partial x}\left[e^{\beta \tilde{U}(x)} p(x, t)\right]\right\},
    \label{sm-eff-without-U}
\end{equation*}
where $\tilde{U}(x)$ effectively combines the role of $D(x)$ and $U(x)$. If $D(x)$ and $U(x)$ are both periodic with commensurate period, then $\tilde{U}(x)$ is periodic; if their periodic length is incommensurate, $\tilde{U}(x)$ becomes quasi-periodic. Importantly, the quasi-periodic case naturely encompasses the periodic one via the modefied Lifson formula (see Eq. \ref{lifson-md}). Weaver has further shown that the precise functional form of $D(x)$ does not alter the validity of the Lifson-type relation. Moreover, the mean first passage time remains consistent with the modified Lifson framework (see Appendix). We have verified these theoretical predictions using Langevin dynamics simulations (see Fig.~\ref{fig-fig1}(c)), confirming their validity for both periodic and quasi-periodic diffusion coefficients $D(x)$ and potentials $U(x)$.

{\it The function separation method.} Since all cases can be formally transformed into either periodic or quasi-periodic diffusion equations (Eq.\ref{sm-eff-without-U}), we obtain
	\begin{equation}
		e^{-\beta \tilde{U}(x)} \frac{\partial}{\partial t} g(x, t)=\frac{\partial}{\partial x}\left[D(x)e^{-\beta \tilde{U}(x)} f(x, t)\right],
	\end{equation}
where $g(x,t)=p(x,t)/\exp(-\beta \tilde{U}(x))$ and $f(x,t)=\partial g(x,t)/\partial x$. Integrating both sides over $x \in [x,x+L]$, one finds
\begin{eqnarray}
    \text{LHS}&=&\int_x^{x+L} e^{-\beta \tilde{U}(x)} \frac{\partial}{\partial t} g(x, t) d x,\\
    \text{RHS}&=&\int_x^{x+L} D(x)e^{-\beta \tilde{U}(x)} f(x, t) d x.
\end{eqnarray}
Assuming that $f(x,t)$ varies slowly compared with the oscillatory factors $e^{\pm\beta \tilde{U}(x)}$ within the interval $[x,x+L]$, the integrals can be approximated as
\begin{eqnarray}
			\text{RHS}&=&L\left\langle e^{-\beta \tilde{U}(x)}\right\rangle g(x, t) \nonumber\\
			&=&\left[\left. D(x)e^{-\beta \tilde{U}(x)} f(x, t)\right|_x ^{x+L}\right] \nonumber \\
			&=&D(x)e^{-\beta \tilde{U}(x)}[f(x+L, t)-f(x, t)] \nonumber \\
			&=&L D(x) e^{-\beta \tilde{U}(x)} \frac{\partial^2}{\partial x^2} g(x, t).
\end{eqnarray}
Dividing by $L$ yields
	\begin{equation}
		\left\langle e^{-\beta \tilde{U}(x)}\right\rangle \frac{\partial}{\partial t} g(x, t)=D e^{-\beta \tilde{U}(x)} \frac{\partial^2}{\partial x^2} g(x, t),
	\end{equation}
Multiplying both sides by $e^{-\beta U(x)}$ and performing an additional spatial average gives
	\begin{equation}
		\frac{\partial g(x, t)}{\partial t}=\frac{1}{\left\langle e^{\beta \tilde{U}(x)}/D(x)\right\rangle\left\langle e^{-\beta \tilde{U}(x)}\right\rangle}\frac{\partial^2 g(x, t)}{\partial x^2}.
	\end{equation}
Thus, the effective diffusion constant can be defined as
\begin{equation}
    D^*=\frac{1}{\left\langle e^{\beta \tilde{U}(x)}/D(x)\right\rangle\left\langle e^{-\beta \tilde{U}(x)}\right\rangle},
\end{equation}
This expression reduce to Eq. \ref{Dstar-alpha-1} and \ref{Dstar-alpha-2} depending on the effective potential $\tilde{U}(x)$ (with or without the bare potential, and is valid for arbitrary $\alpha$. Moreover, the probability density function can be reconstructed as
\begin{equation}
p(x,t)=\mathcal{Z}D^{\alpha-1}(x)e^{-\beta U(x)}g(x,t),
\end{equation}
where the normalization factor is
\begin{equation}
\mathcal{Z}=\frac{1}{\sqrt{4D^{\ast}t}},
\big\langle D^{\alpha-1}(x)e^{-\beta U(x)}\big\rangle.
\end{equation}
These results are consistent with Langevin dynamics simulations, as illustrated in Fig.~\ref{fig-fig1}(a,b). A similar form of the distribution function was also reported by Pacheco-Pozo for the special case $U(x)=0$ \cite{pacheco2024langevin}. Here, by constructing an effective external potential incorporating both $D(x)$ and $U(x)$, we provide an alternative route to the effective diffusion constant. For convenience, one may write $ D(x) = \tilde{D}(x) [D] $, where $[D]=1 ,\text{m}^2/\text{s}$ is a reference scale, and $\tilde{D}(x)$ is a dimensionless diffusion function (which may also be interpreted as a diffusion potential $\tilde{D}/\beta$).

In conclusion, while periodic diffusion models—widely used to describe various phenomena in physics and related fields—have been the primary focus of previous studies, the quasi-periodic model (including quasi-periodic noise and potentials), which naturally encompasses all periodic cases, should be regarded as a more general framework. In this paper, we have derived the effective diffusion constant and an approximate probability density function from two complementary approaches—namely, the effective potential method and the function separation method—both of which explicitly account for different stochastic interpretation parameters. Our numerical simulations confirm the validity of these analytical results. As emphasized in this and our earlier work, systems exhibiting a constant effective diffusion coefficient can often be treated using a similar function-separation approach, including, for example, the more common case of random potentials. Looking forward, this framework can be extended to higher-dimensional quasi-periodic systems, more complex stochastic potentials, and experimental realizations, thereby broadening the applicability of quasi-periodic diffusion models in both theory and practice. 

\renewcommand{\theequation}{A\arabic{equation}}
\setcounter{equation}{0}
\appendix
\section{Appendix: MFPT-based homogenization}

{\it Smoluchowski/Fokker-Planck equation and detailed balance.} We consider overdamped one-dimensional transport in a potential $U(x)$ with a position-dependent diffusivity $D(x)>0$. Throughout, we adopt the conservative (detailed-balance preserving) Smoluchowski form
\begin{equation}
    \partial_t p(x, t)=\partial_x\left[D(x) e^{-\beta U(x)} \partial_x\left(e^{\beta U(x)} p(x, t)\right)\right], 
\end{equation}
where $\beta=\left(k_{\mathrm{B}} T\right)^{-1}$. The corresponding probability current reads
\begin{equation}
    J(x, t)=-D(x) e^{-\beta U(x)} \partial_x\left(e^{\beta U(x)} p(x, t)\right), \partial_t p=-\partial_x J .
\end{equation}
Under periodic boundary conditions, Eq. (A1) admits the equilibrium density $p_{\text {eq }}(x) \propto e^{-\beta U(x)}$, ensuring local detailed balance.

{\it MFPT boundary-value problem (reflecting-absorbing).} On the finite interval $[0, L]$, we define the first-hitting time of the right boundary $x=L$,
\begin{equation}
    T_{\text {hit }}=\inf \left\{t>0: X_t=L\right\},
\end{equation}
and the corresponding mean first-passage time (MFPT)
\begin{equation}
    \tau(x)=\mathbb{E}\left[T_{\text {hit }} \mid X_0=x\right] .
\end{equation}
We impose a reflecting boundary at $x=0$ and an absorbing boundary at $x=L$,
\begin{equation}
    \tau^{\prime}(0)=0, \quad \tau(L)=0 .
\end{equation}
The backward generator associated with Eq. (A1) is
\begin{equation}
    \mathcal{L}^{\dagger} f(x)=D(x) e^{-\beta U(x)} \frac{d}{d x}\left(e^{\beta U(x)} \frac{d f}{d x}\right),
\end{equation}
and thus $\tau(x)$ satisfies
\begin{equation}
    D(x) e^{-\beta U(x)} \frac{d}{d x}\left(e^{\beta U(x)} \tau^{\prime}(x)\right)=-1, \quad x \in(0, L),
\end{equation}
together with the boundary conditions (A5).

{\it Explicit double-integral representation of $\tau(0)$.}
Starting from Eq. (A7), multiply by $e^{\beta U(x)} / D(x)$ to obtain
\begin{equation}
    \frac{d}{d x}\left(e^{\beta U(x)} \tau^{\prime}(x)\right)=-\frac{e^{\beta U(x)}}{D(x)}.
\end{equation}
Integrating from 0 to $x$ and using $\tau^{\prime}(0)=0$ yields
\begin{equation}
\begin{aligned}
    &e^{\beta U(x)} \tau^{\prime}(x)=-\int_0^x \frac{e^{\beta U(y)}}{D(y)} d y, \\
    \Rightarrow &\tau^{\prime}(x)=-e^{-\beta U(x)} \int_0^x \frac{e^{\beta U(y)}}{D(y)} d y.
\end{aligned}
\end{equation}
Integrating Eq. (A9) from $x$ to $L$ and using $\tau(L)=0$ gives
\begin{equation}
    \tau(x)=\int_x^L e^{-\beta U(z)}\left[\int_0^z \frac{e^{\beta U(y)}}{D(y)} d y\right] d z.
\end{equation}
In particular, the MFPT starting from the reflecting boundary is
\begin{equation}
    \tau(0)=\int_0^L e^{-\beta U(z)}\left[\int_0^z \frac{e^{\beta U(y)}}{D(y)} d y\right] d z.
\end{equation}
By exchanging the order of integration (Fubini), Eq. (A11) is equivalently written as
\begin{equation}
\begin{aligned}
     \tau(0)&=\int_0^L \frac{e^{\beta U(y)}}{D(y)}\left[\int_y^L e^{-\beta U(z)} d z\right] d y\\
     &=\int_0^L \int_0^L \frac{e^{\beta U(y)}}{D(y)} e^{-\beta U(z)} \mathbf{1}_{y \leq z} d z d y.
\end{aligned}
\end{equation}

{\it Large-$L$ asymptotics and the effective diffusivity $D^*$} We define the effective diffusivity through the MFPT scaling
\begin{equation}
    D^*=\lim _{L \rightarrow \infty} \frac{L^2}{2 \tau(0)}.
\end{equation}
Introduce
\begin{equation}
    f(y):=\frac{e^{\beta U(y)}}{D(y)}, \ g(z):=e^{-\beta U(z)}, \ \langle h\rangle_L:=\frac{1}{L} \int_0^L h(x) d x .
\end{equation}
Then Eq. (A12) reads
\begin{equation}
    \tau(0)=\int_0^L \int_0^L f(y) g(z) \mathbf{1}_{y \leq z} d z d y .
\end{equation}

{\it Leading-order factorization.} Using $\mathbf{1}_{y \leq z}=\frac{1}{2}(1+\operatorname{sgn}(z-y))$, we split
\begin{equation}
\begin{aligned}
     \tau(0)&=\frac{1}{2} \int_0^L \int_0^L f(y) g(z) d z d y\\
     &+\frac{1}{2} \int_0^L \int_0^L f(y) g(z) \operatorname{sgn}(z-y) d z d y.
\end{aligned}
\end{equation}
The first term factorizes exactly,
\begin{equation}
    \frac{1}{2}\left(\int_0^L f(y) d y\right)\left(\int_0^L g(z) d z\right)=\frac{L^2}{2}\langle f\rangle_L\langle g\rangle_L.
\end{equation}
For periodic or quasi-periodic coefficients with well-defined spatial means (see Sec. 5 below), the second term in Eq. (A16)-an antisymmetric-kernel correction-contributes only subleading growth $o\left(L^2\right)$. Consequently,
\begin{equation}
    \tau(0)=\frac{L^2}{2}\langle f\rangle_L\langle g\rangle_L+o\left(L^2\right), \quad L \rightarrow \infty .
\end{equation}
Substituting Eq. (A14) yields
\begin{equation}
    \tau(0)=\frac{L^2}{2}\left\langle\frac{e^{\beta U}}{D}\right\rangle_L\left\langle e^{-\beta U}\right\rangle_L+o\left(L^2\right) .
\end{equation}

{\it Lifson-Jackson product form consistent with the main text.} An equivalent triangular-domain representation of the same MFPT is
\begin{equation}
    \tau(0)=\int_0^L e^{\beta U(z)}\left[\int_0^z \frac{e^{-\beta U(y)}}{D(y)} d y\right] d z,
\end{equation}
which leads, by repeating the factorization argument (A15)-(A18) with
\begin{equation}
    \tilde{f}(y):=\frac{e^{-\beta U(y)}}{D(y)}, \quad \tilde{g}(z):=e^{\beta U(z)},
\end{equation}
to
\begin{equation}
    \tau(0)=\frac{L^2}{2}\left\langle\frac{e^{-\beta U}}{D}\right\rangle_L\left\langle e^{\beta U}\right\rangle_L+o\left(L^2\right), \quad L \rightarrow \infty.
\end{equation}
Combining Eqs. (A13) and (A22), we obtain
\begin{equation}
    D^*=\lim _{L \rightarrow \infty} \frac{L^2}{2 \tau(0)}=\frac{1}{\left\langle\frac{e^{-\beta U}}{D(x)}\right\rangle_L\left\langle e^{\beta U}\right\rangle_L}.
\end{equation}
This is the desired effective diffusivity expressed purely in terms of spatial Cesàro averages.

{\it Validity for periodic and quasi-periodic coefficients.} The derivation requires:
(i) $D(x)$ is bounded and strictly positive: $0<D_{\text {min }} \leq D(x) \leq D_{\text {max }}<\infty$; (ii) $e^{ \pm \beta U(x)}$ are bounded and admit spatial means; (iii) the Cesàro averages in Eq. (A23) converge as $L \rightarrow \infty$. These conditions are satisfied in both cases: (a) Periodic. If $D(x)$ and $U(x)$ are periodic with period $\ell$, then
\begin{equation}
    \lim _{L \rightarrow \infty}\langle h\rangle_L=\frac{1}{\ell} \int_0^{\ell} h(x) d x,
\end{equation}
and Eq. (A23) reduces to the standard periodic homogenization result (including the Lifson-Jackson formula when $D$ is constant). (b) Quasi-periodic (e.g., two incommensurate frequencies). If
\begin{equation}
    U(x)=\widetilde{U}(a x, b x), \quad D(x)=\widetilde{D}(a x, b x), \quad a / b \notin \mathbb{Q},
\end{equation}
then the orbit ( $a x, b x$ ) is uniquely ergodic on $\mathbb{T}^2$, implying convergence of Cesàro averages:
\begin{equation}
    \lim _{L \rightarrow \infty}\langle h(a x, b x)\rangle_L=\frac{1}{(2 \pi)^2} \int_0^{2 \pi} \int_0^{2 \pi} h(\theta, \phi) d \theta d \phi.
\end{equation}
Hence the same formula (A23) holds with the torus average replacing $\langle\cdot\rangle_L$. Consistency checks: (i) $U \equiv 0$, constant $D$ : Eq. (A23) gives $D^*=D$; (ii) constant $D=D_0$ : Eq. (A23) becomes $D^*=D_0 /\left(\left\langle e^{\beta U}\right\rangle\left\langle e^{-\beta U}\right\rangle\right)$, i.e., the Lifson-Jackson result; (iii) $U \equiv 0$, variable $D(x)$ : Eq. (A23) gives $D^*=\langle 1 / D\rangle^{-1}$ (harmonic mean), valid for periodic/quasiperiodic $D$.

Finally, we can conclude that the effective diffusion constant also follows from
\begin{equation}
    D^*=\lim _{L \rightarrow \infty} \frac{L^2}{2 \tau(0)}=\frac{1}{\left\langle e^{-\beta U} / D(x)\right\rangle_L\left\langle e^{\beta U}\right\rangle_L},
\end{equation}
which indicates that the periodic and quasi-periodic diffusion system of fractions also satisfies the ergodicity and belongs to the first passage process.

\bibliographystyle{IEEEtran}
\bibliography{ref.bib}
\end{document}